# Single-Shot Time-Domain Studies of Spin-Torque-Driven Switching in Magnetic Tunnel Junctions

Y.-T. Cui[1], G. Finocchio[2], C. Wang[1], J. A. Katine[3], R. A. Buhrman[1] and D. C. Ralph[1]

[1]*Cornell University, Ithaca, NY 14853, USA*

[2]*Dipartimento di Fisica della Materia e Ingegneria Elettronica, University of Messina, Salita Sperone 31, 98166 Messina, Italy*

[3]*Hitachi Global Storage Technologies, San Jose, CA 95135 USA*

We report single-shot measurements of resistance versus time for thermally assisted spin-torque-driven switching in magnetic tunnel junctions. We achieve sufficient sensitivity to resolve the resistance signals leading up to switching, including the variations between individual switching events. Analyses of pre-switching thermal fluctuations allow detailed measurements of coherence times and variations in magnetization precession amplitude. We find that with a small in-plane hard-axis magnetic field the magnetization dynamics are more spatially coherent than for the case of zero field.

Magnetization switching induced by spin transfer torque [1,2] is of interest both for probing the fundamental physics of magnetic dynamics and for applications in storage technologies [3,4]. Measurements in the time domain [5-11] can provide the most direct information about spin-torque-driven magnetic dynamics. However, most previous time-resolved studies [5-9] employed stroboscopic approaches, which average over many events so that they reveal only average behavior and hide individual variations. Magnetic tunnel junctions (MTJs) can now provide sufficiently large resistance signals (relative to metal spin valves [5-8]) for single-shot measurements. Two initial experiments have measured distributions of spin-torque switching times in MTJs using single-shot techniques [10,11] but they did not resolve the dynamics leading to switching. Here we report single-shot measurements of spin-torque switching in MTJs with sufficiently improved sensitivity to study the pre-switching resistance signals in detail. We observe the variations between switching processes caused by thermal fluctuations and can perform comprehensive analyses of the fluctuations prior to switching. We find that switching is more spatially coherent when the magnetic moments of the electrodes are initially offset (at an angle different than 0° or 180°) than when the moments are collinear.

The MTJ samples that we study have the layer structure (in nm): bottom contact [{Ta(3)/CuN(41.8)}$_2$/Ta(3)/Ru(3.1)], synthetic antiferromagnet (SAF) pinned layer [IrMn(6.1)/CoFe(1.8)/Ru/CoFeB(2.0)], tunnel barrier [MgO], magnetic free layer [CoFe(0.5)/CoFeB(3.4)], capping layer [Ru(6)/Ta(3)/Ru(4)]. The resistance-area product is 3 Ω μm$^2$ and the CoFe/CoFeB free layer that undergoes switching has a saturation magnetization-thickness product $M_s t = 3.2 \times 10^{-4}$ emu/cm$^2$. Both the pinned and free layers are patterned into an approximately elliptical cross section of 65 × 130 nm$^2$ with long axis parallel to the pinning of SAF layer. All measurements are performed at room temperature, and positive current corresponds to electron flow from the free to the pinned layer. A hysteresis loop for a magnetic field applied along the easy axis is shown in Fig. 1(a), indicating an effective field on the free layer $H_d$ = 60 Oe and an in-plane anisotropy $H_K \approx$ 280 Oe, after accounting for thermal fluctuations [12]. Our discussion of current-driven reversal will focus on switching from the anti-parallel (AP) state to parallel (P) state because this required approximately 30% lower voltages than P-to-AP switching.

However, measurements of P-to-AP dynamics are qualitatively similar. We will examine switching both in the case of zero total field on the free layer (in which an applied field cancels $H_d$) and the case that a small (100 Oe) field is applied along the in-plane hard axis, to rotate the free-layer magnetization approximately 15° from the strictly AP configuration (see inset, Fig. 1(a)). We will present data from a single sample, but have studied two other devices in zero field and one other in the hard-axis field, with all showing the same differences depending on field geometry.

We perform single-shot measurements using the circuit shown in Fig. 1(b). After initializing the sample in the AP state we use a pulse generator to produce an 100 ns long, 300 ps rise-time negative pulse, split this signal with a power divider, and apply one part with amplitude $V_{\text{inc}}$ to the sample through one high-frequency probe. We detect the transmitted pulse with a second probe, and amplify initially with one stage of an inverting amplifier with effective gain of 5 dB. A key feature of the measurement is that we then combine the transmitted signal with the split-off copy of the original pulse using a time delay and attenuation factor tuned so as to cancel the part of the transmitted pulse that does not depend on the magnetic dynamics. This improves our dynamic range, enabling us to detect small resistance changes, because we can apply additional amplification without saturating the amplifiers before or during the pulse. We record the signal with an 11 GHz, 40 GSamples/s storage oscilloscope. When plotting AP-to-P switching events, we subtract an average baseline curve taken with the sample initialized to the P state, for which the negative current pulse does not produce switching. The transmitted pulse at the sample (before amplification) is $V(t) = V_{\text{inc}}/[1 + G_S(t)Z_0/2]$, where $G_S(t)$ is the sample conductance (the reciprocal of the resistance) and $Z_0 = 50\ \Omega$ is the probe impedance. For cases that the conductance change $\Delta G_S(t) = G_S(t) - G_P$ is much less than the parallel conductance $G_P$, we then have after amplification that the measured signal is

$$\Delta V_{\text{meas}}(t) = -\frac{G_{\text{amp}} Z_0}{2(1 + (Z_0/2)G_P)^2} \Delta G_S(t) V_{\text{inc}}, \tag{1}$$

with $G_{\text{amp}} = 14$ dB the total amplification. Figure 1(c) shows a representative trace for AP-to-P reversal, for the 100 Oe hard-axis field and a transmitted pulse amplitude $V = -750$ mV. We have the sensitivity to observe conductance variations above the noise background throughout the period between the pulse onset and the switching event. The rms noise level corresponds to ~ 2% of the difference between AP and P conductances.

Before examining the time traces in detail, we will first discuss the statistics of the switching times, to demonstrate that our data correspond to thermally-assisted spin-torque switching [13,14]. Figure 1(d) shows the switching-time distributions for three values of $V$, for the case of the 100 Oe hard-axis applied field. The broad widths of the distributions indicate that our pulse amplitudes |V| are below the zero-temperature switching threshold so that switching is assisted by thermal fluctuations. The long-time tails of the distributions fit well to a simple exponential $\exp[-t/\tau_0(V)]$ expected for thermally activated switching. Extrapolating the measured values of $\tau_0(V)$ (Fig. 1(d), inset) to 1 ns gives an estimate for the zero-temperature switching threshold of -870 mV. We have not attempted to investigate higher values of |V| because the MTJs are not sufficiently stable. The distributions in Fig. 1(d) depart from the exponential form with a peak at short times (ranging from $t = 4$ ns for $V = $ -750 mV to 20 ns for $V = $ -600 mV), indicating that time is required after the pulse onset before the fluctuations reach an effective equilibrium [11].

Now we turn to the details of individual traces. In Fig. 2(a) and (b), we show selected traces with different switching times for $V = $ -750 mV, for both the 100 Oe in-plane hard-axis field and zero total field. We resolve the resistance oscillations prior to switching in both configurations, with oscillation amplitudes that fluctuate with time. In the right columns of Fig. 2(a,b) we zoom in to each trace 2 ns before and after switching. For the 100 Oe hard-axis field (Fig. 2(a)), most traces show at least a few cycles of large oscillations in the 2 ns prior to switching, corresponding to magnetic precession with large amplitude leading to reversal [1,3,6]. However, these oscillations show large variations from trace to trace, ranging from almost-vanishing amplitude (2$^{nd}$ curve in Fig. 2(a)) to oscillations close to one half the difference between the initial ($\approx$ AP) and final ($\approx$ P) values. For the zero total field case, the resistance oscillations immediately prior to switching are much weaker (Fig. 2(b)). Many traces for this case merely show a gradual increase in $\Delta V_{meas}$ without any significant oscillations immediately prior to switching.

Figures 2(c,d) show averages of 2000 traces with their switching edges aligned, for the two field geometries. We first note that these averaged curves are very different from typical individual traces. Still, for the 100 Oe hard-axis field the averaged trace shows oscillations with increasing amplitude for a few cycles before switching (Fig. 2(c)), indicating that switching occurs preferentially at a particular phase of the resistance oscillations. The (1/$e$) time scale for the averaged precession amplitude to build prior to reversal is 0.25 ± 0.02 ns for all $V$ from -540 mV to -750 mV. This is likely a measure of the correlation time for dephasing between different traces with differing amplitudes of large-angle precession, rather than a true measure of precession amplitude changes, because this scale is shorter than the coherence time for amplitude changes within individual traces (determined below). For the averaged trace in the case of zero total field (Fig. 2(d)), the oscillatory features are almost entirely washed out, suggesting much weaker correlations between the oscillation phase and the switching time.

We have performed micromagnetic simulations to understand these results, using the code described in references [15,16]. The simulation parameters are: free layer saturation magnetization $M_S$ = 1050 emu/cm$^3$, damping = 0.025, exchange = 1.3 × 10$^{-11}$ J/m, uniaxial anisotropy 4 × 10$^3$ J/m$^3$, spin polarization = 0.6, and sample temperature = 400 K. The sample size is the same as in the experiment, the current density during the pulse is -2 × 10$^6$ A/cm$^2$, and the pinned layer is assumed to be immobile. In Fig. 3, we plot examples of simulated conductance traces for the two field configurations discussed above and also the averaged traces over 100 simulated reversals near the switching edges (compare to Figs. 2(c) and (d)). The simulations reproduce many of the features seen in the experiment including the differences in the coherence of the oscillations between the two field geometries. The calculated magnetic configurations during switching events (Fig. 3(e) and (f)) suggest that the typical mechanism for reversal differs for the two field geometries. Switching for zero total field generally proceeds with one end of the sample switching first and the rest of the sample following by domain wall propagation [10] (Fig. 3(e)). For the hard-axis field, the switching dynamics are generally more spatially uniform, albeit with local perturbations due to thermal fluctuations (Fig. 3(f)).

The difference in the degree of spatial uniformity in the two field configurations can also be observed directly in the time traces long before switching. For zero total field, the local minima in the fluctuations of $\Delta V_{meas}(t)$ exhibit excursions from the global minimum on scales of several ns, much longer than any precessional period (see especially the 3$^{rd}$ and 4$^{th}$ panels in Fig. 2(b)). For the 100 Oe hard-axis field, the local minima in $\Delta V_{meas}(t)$ are more closely clustered near the AP value (Fig. 2(a)). Large excursions of the sort present for the zero-field case are inconsistent with any approximately spatially-homogeneous dynamics for a thin-film magnetic sample, since at zero field the free-layer moment should pass close to the AP configuration twice per precessional cycle, each time giving the same global minimum in $\Delta V_{meas}(t)$.

The excellent sensitivity of our measurements allows us to analyze the statistical properties of the magnetic dynamics even where the conductance oscillations are small, well away from the switching event. Fig. 4(a) plots autocorrelation functions of the conductance versus time for the full interval between pulse onset and switching in the case of the 100 Oe hard axis field, averaged over all switching traces with switching times longer than 10 ns at each value of $V$ and normalized to the full difference between initial and final conductances. We find decoherence times $\Delta t_c$ ranging from 0.54 ns at -540 mV to 0.45 ns at -750 mV (inset, Fig. 4(b)). To help distinguish between frequency and amplitude variations, we plot in Fig. 4(b) autocorrelation functions of the oscillation amplitude versus peak number, determined by measuring every conductance peak between the pulse onset and switching. The coherence times for amplitude fluctuations $\Delta t_A$ are greater than but comparable to $\Delta t_c$ at each value of $V$, from which we conclude that both frequency and amplitude variations are significant in these ≥ room temperature fluctuations. Surprisingly, the coherence times decrease with increasing values of |$V$|, whereas spin torque for this bias should decrease the effective damping and should therefore increase the coherence time. We interpret the decrease in the coherence times with |$V$| to be a sign of heating in these nonlinear oscillators. Theories of magnetic nano-oscillators have not yet considered the combined effects of temperature and nonlinearities in the thermally-activated switching regime [17], but we speculate that a

Page **6** of 13

more detailed analysis of fluctuations and coherence times may provide a direct measure of magnetic heating in the precessing layer.

We can also Fourier transform our time traces for the pre-switching dynamics to achieve a measurement equivalent to thermally-excited ferromagnetic resonance [18], but accomplished for the short-lived nonequilibrium state before switching. Figures 4(c) and (d) show power spectra for the interval between the pulse onset and switching, averaged over all traces with switching times longer than 10 ns at each value of $V$ for the two field geometries. For the 100 Oe hard-axis field (Fig. 4(c)), the spectra show a well-defined peak which with increasing $|V|$ grows in amplitude and shifts to lower frequency. These results can be understood in terms of approximately spatially-uniform dynamics. The increasing amplitude of the resonance can be explained by the reduction in effective magnetic damping due to the spin-transfer torque [1,3], together with heating. The frequency shift can be understood as due primarily to the dependence of frequency on precession amplitude [19]. We estimate that the rms precesion angle ranges from 8º for $V$ = -540 mV to 14º for $V$ = -750 mV for the 100 Oe hard-axis field. For zero total field (Fig 4(d)), the average Fourier spectra show much weaker peaks, and possess a low-frequency tail. This is another indication of incoherent dynamics for this field geometry.

In summary, we have performed single-shot measurements of the conductance during thermally-assisted spin-torque-induced switching in MTJs. The measurements provide a detailed view of the switching dynamics, and how they vary between switching events. We observe that switching is more spatially homogeneous when the magnetic moments of the electrodes are initially offset by a small hard-axis magnetic field, compared to an initial collinear configuration. Our measurements also allow for detailed analyses of the nonequilibrium magnetic fluctuations preceding switching.

We thank P. Brouwer, B. Azzerboni, and L. Torres for discussions and D. Mauri for growing the MTJ layers. This work is supported by ONR, ARO, NSF (DMR-0605742), Spanish project MAT2008-04706/NAN, and NSF/NSEC through the Cornell Center for Nanoscale Systems. We also acknowledge NSF support through use of the Cornell Nanofabrication Facility/NNIN and the Cornell Center for Materials Research facilities.

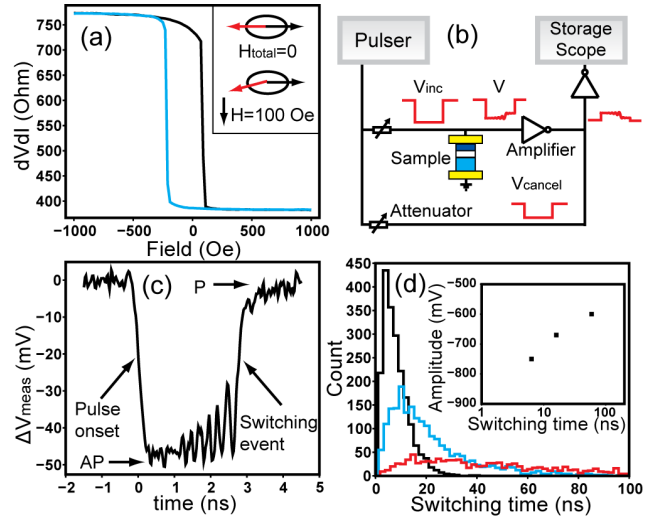

Fig. 1. (color online) (a) Resistance vs. magnetic field applied along the easy axis. Inset: initial magnetization configurations for zero total field and for an 100 Oe in-plane hard-axis field for which the free layer rotates by ~ 15º relative to the stationary fixed layer. (b) Schematic of the measurement circuit. (c) An example of an AP-to-P switching trace after baseline subtraction for $V$ = -750 mV and the 100 Oe hard-axis field. (d) Histograms of switching times for $V$ = -750 mV (narrowest distribution), -670 mV, and -600 mV (broadest distribution). (inset) Pulse amplitude vs. average switching time.

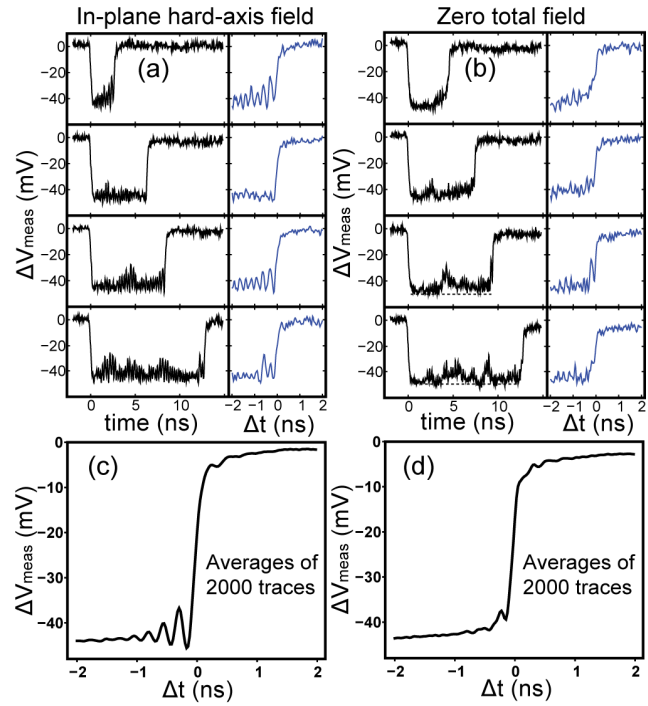

Fig. 2. (color online) Measured oscillatory signals leading up switching for $V = -750$ mV. (a),(b) Representative switching traces for (a) an in-plane hard-axis field of 100 Oe and (b) zero total field, together with magnified views 2 ns before and 2 ns after the switching events. (c),(d) Average of 2000 measured traces with the switching edge aligned for (c) the in-plane hard-axis field and (d) zero total field.

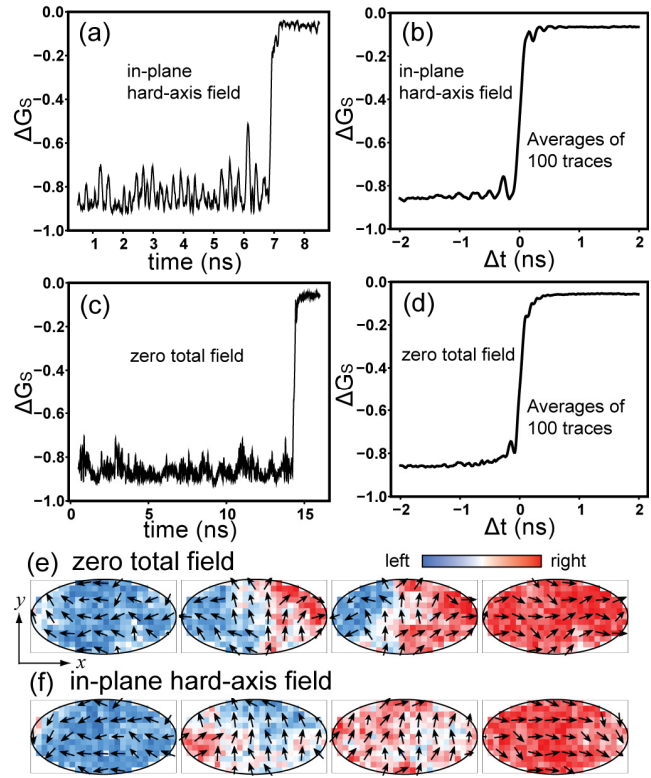

Fig. 3. (color online) (a),(c) Simulated switching traces for the two field geometries. The change in conductance is normalized by the difference $G_P$ - $G_{AP}$. (b),(d) Averages of 100 simulated switching traces with the switching edges aligned. (e),(f) Snapshots with 85 ps spacing of micromagnetic configurations during the switching events. The color scale denotes the magnetization component along the x axis.

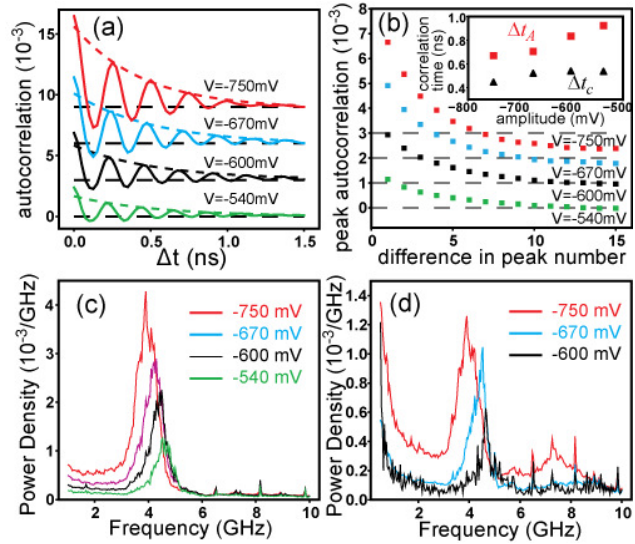

Fig. 4. (color online) (a) Autocorrelation functions of pre-switching conductance variations, normalized to the conductance difference before and after switching. The dotted lines are exponential fits to the peaks. The curves are offset vertically with horizontal lines denoting zero. (b) Autocorrelation functions for the conductance peak amplitude versus peak number. Inset: correlation times $\Delta t_c$ from (a) and $\Delta t_A$ from (b). Both (a) and (b) correspond to the 100 Oe in-plane hard-axis field configuration. (c) Averaged Fourier spectra of normalized conductance variations for the hard-axis field configuration. (d) Averaged Fourier spectra for zero total field.